\documentclass[article,showpacs,amsmath,amssymb,aps]{revtex4}
\usepackage{graphicx}
\usepackage{epsfig}
\usepackage{color}
\usepackage{xcolor}
\begin{document}
\title{Semi-classical and boson descriptions of scissors states}

\author{A. A. Raduta$^{a),b)}$,  C. M. Raduta$^{a)}$ and Robert Poenaru$^{c)}$ and Al. H. Raduta$^{a)}$}

\address{$^{a)}$ Department of Theoretical Physics, Institute of Physics and
  Nuclear Engineering, Bucharest, POBox MG6, Romania}

\address{$^{b)}$Academy of Romanian Scientists, 54 Splaiul Independentei, Bucharest 050094, Romania}

\address{$^{c)}$Department of Inovation, Orange Services, Str. Gara Herastrau 4D, Bucharest, Romania} 

\begin{abstract}
A two interacting rotors Hamiltonian is alternatively treated  semi-classically and by a Dyson boson expansion method. The linearized equations of motion lead to dispersion equation for the wobbling frequency. One defined a ground band with energies consisting in a rotational part and one half of the vibrational wobbling energy. Adding to each state energy the corresponding wobbling quanta one obtains the first excited band. Phonon amplitudes are used to calculate the reduced probability for the inter-band  M1 transitions. The states  exhibit a shears character. One points out a chiral symmetry which is broken by the interaction term, leading to a  pair of twin chiral bands. Applications are made for $^{156}$Gd. One outlines the ability of the two rotor model to account for the wobbling and chiral motion in nuclei. Although the chosen trial function has not a definite total angular momentum, for two particular ansatz of the pairs $I_p,I_n$ the average value of the total angular momentum approximates,
 to a certain accuracy, the partial angular momentum $I_p$ In this context, the rotational bands defined throughout this present paper  could be labeled by the total I.
\end{abstract}
\maketitle
corresponding author: A. A. Raduta

email address: raduta@nipne.ro

\renewcommand{\theequation}{1.\arabic{equation}}
\setcounter{equation}{0}
\section{Introduction}
The description of magnetic properties in nuclei has always been a
central issue. The reason is that the two systems of protons and
neutrons respond differently when they interact with an external
electromagnetic field. Differences are due to the fact that by
contrast to neutrons, protons are charged particles, the proton and
neutron magnetic moments are different from each other and finally
the protons and neutron numbers are also different.  In 1965
 Greiner and his collaborators noticed that using different
moments of inertia for proton and neutron systems \cite{GR64} 
has notable impact on energies and magnetic transitions. Later on the mentioned authors
elaborated  the two liquid drops
model aimed at  describing the isovector $2^{+}$ state
\cite{FA66} as well as the M1 properties \cite{MGM75} of the
rotational bands. The first microscopic description for the
magnetic dipole states was proposed by Gabrakov, Kuliev and Pyatov in Ref.
\cite{GKP70},where the single particle motion was described by a deformed Woods Saxon mean field.  The same authors succeeded to
eliminate the  spurious contributions to the magnetic mode
\cite{KP74}. Rowe also studied the magnetic dipole mode within potential vibrating method \cite{SR77}. However the group which
brought something essentially new in this field is that of Lo Iudice
and Palumbo \cite{IPF84} who proposed a phenomenological model,
called Two Rotor Model (TRM), which assigns to the proton and
neutron systems two axially symmetric rigid rotors, having
different symmetry axes. The mode appears to be determined by angle vibrations 
of the two symmetry axes. This picture inspired
the naming as ``scissors mode''. Within the TRM, the mode  
is excited due to the interaction of
the nuclear convection current with the electromagnetic field.
Despite the fact that the predictions of the TRM for both energy and M1
probability to be excited, are much larger than the experimental
data, obtained few years later, the big merit of this model is to
predict a pure orbital mode of collective nature, without
involving the spin degrees of freedom. This
feature was confirmed by all microscopic calculations. The field of
collective M1 states was enormously stimulated by the group of
Richter, which identified the M1 state, for $^{156}$Gd, in a high
resolution $(\text{e},\text{e}')$ experiment at backward angle
\cite{BRS83}. The results for  excitation energy and  B(M1)
value were confirmed by a nuclear resonance fluorescent experiment
\cite{BBD84}. Since then, many experiments have been performed and
the number of nuclei known to exhibit a scissors mode  was enlarged
by many rear earth and actinides nuclei but also by some medium
isotopes from the Ti region. Another phenomenological model aimed at
describing the measured properties of $1^{+}$ is the interacting
boson model (IBA2). In this model, the M1 state is caused by
breaking the F spin symmetry by a Majorana interaction of the
proton-like and neutron-like bosons. The state energy is obtained by
a suitable fixing of the interaction strength. Thus IBA2 is not
making predictions for the state energy but only for the M1
excitation probability from the ground state \cite{IA81}.
The generalized coherent state model (GCSM) \cite{RFC87} has been used to describe simultaneously the scissors mode and the first three major collective bands, ground, $\beta$ and $\gamma$. The important finding of the proposed model is that the total M1 strength is proportional to the nuclear deformation squared, which proves the collective nature of the mode \cite{IRD94}.
Two scissors modes were defined, one ($1^+$) being a two  proton-neutron phonon excitation of a deformed ground state
while the other one ($\tilde{1}^+)$ is an isovector quadrupole boson excitation of the deformed state heading the beta band.
The merit of the GCSM is that on top of each of the mentioned dipole scissors states one builds up a full rotational band. The band $1^+$ is formed of odd spin states while
the band $\tilde {1}^+$ band is a $\Delta I=1$ band exhibiting a pronounced staggering structure.
It is worth mentioning that the first predictions for the scissors mode in even-odd nuclei were achieved in Refs. \cite{RI89,RD90}. 

The field of scissors states is still actual. Indeed, in the meantime many sophisticated microscopic descriptions were proposed.Thus, the covariant density functional theory was used to analyze the evolution of the low lying M1 strength in superfluid nuclei in the framework of the Relativistic Quasiparticle Random Phase Approximation (QRPA). In nuclei with large neutron excess two scissors modes were identified, one being the conventional scissors mode, while the second one is a new soft mode describing the scissors-like oscillations of the neutron skin against a deformed proton-neutron core \cite{Pena}.A self-consistent RPA with Skyrme interaction is used in Ref.\cite{Nest} to describe orbital and spin-flip modes in the isotopic chain $^{142-152}$Nd. Only a qualitative 
description of the data is provided. The  Cogny interaction is a specific effective interaction used in  nuclear structure theory particularly within the Skyrme-Hartree-Fock-Bogoliubov framework, which provides a more accurate, beyond-mean-field description of the scissors mode by accounting for the complex short-range forces between nucleons.
An analysis of the accumulated $\gamma$-ray spectra following neutron capture within the extreme statistical model lead to the conclusion that scissors mode resonances are built not only on the ground state but also on excited levels in the product isotopes of Gd \cite{Kroll}.

To obtain more detailed information about the literature devoted to
this subject, we advise the reader to consult the review papers on
this issue \cite{RI90,IUD95,ZAW98}.

Here we study the magnetic properties of scissors-like states within a two rotors model. We present a new view on this issue namely a two-rotor Hamiltonian, described in Section 2, is 
successively treated through a semi-classical and boson expansion procedures. These objectives are touched in sections 3 and 4, respectively. In these sections the wobbling frequencies for  a two rotor system are obtained. In Section 5 the expressions for  the magnetic dipole reduced probabilities are derived. An extensive numerical analysis is presented in Section 6. One defines a ground and an excited wobbling band are defined whose energies are subsequently calculated. Also the corresponding B()M1) values for  exciting states from the second the second band are calculated. The nature of the found states, scissors or shears, is investigated. 
Final conclusions are drown in Section 7. 
\renewcommand{\theequation}{2.\arabic{equation}}
\setcounter{equation}{0}
\section{Two rotors Hamiltonian}
The objective of this paper is to study the properties of the following two interacting  rotors Hamiltonian:
\begin{equation}
H=\frac{\vec{I}^2_p}{2{\cal J}_p}+\frac{\vec{I}^2_n}{2{\cal J}_n}+V\vec{I}_P\cdot\vec{I}_n.
\end{equation}
with $\vec{I}_p$ and $\vec{I}_n$ denoting the angular momenta carried by the proton and neutron system, respectively. The moments of inertia for the two 
systems, which are supposed to be axially deformed ellipsoids, are denoted by ${\cal J}_p$ and ${\cal J}_n$, respectively.

In terms of raising and lowering angular momenta the model Hamiltonian gets the form:

\begin{equation}
H=\frac{\vec{I}_{p}^2}{2{\cal J}_p}+\frac{\vec{I}_{n}^2}{2{\cal J}_n}+\frac{V}{2}\left[(I_{P+}I_{n-}+I_{p-}I_{n+}+2I_{P3}I_{n3}\right],
\label{Has}
\end{equation}
where:
\begin{equation}
I_{p\pm}=I_{p1}\pm iI_{p2},\;\;I_{n\pm}=I_{n1}\pm iI_{n2},
\end{equation}
with $I_{pk}$, $I_{nk}$, k=1,2,3, being the k-th Cartesian components of the vectors $\vec{I}_p, \vec{I}_n$, respectively, while "i" is the imaginary unit.
\renewcommand{\theequation}{3.\arabic{equation}}
\setcounter{equation}{0}
\section{Semi-classical description}
The main spectroscopic properties of $H$ will be evidenced by  the Time Dependent Variational Principle: 
\begin{equation}
\delta\int\langle \Psi^{I_pI_n}_{M_pM_n}|(H-i\frac{\partial}{\partial t'})|\Psi^{I_pI_n}_{M_pM_n}\rangle dt'=0.
\label{TDVP}
\end{equation}
If the trial function spans the whole Hilbert space generated by the eigenfunction of $H$, then  solving the variational equation
is equivalent with solving the time dependent Schr\"{o}dinger equation. Clearly $H$ commutes with $\vec{I}_{p}^2$ and $\vec{I}_{n}^2$ but not with
$I_{p3}$ and $I_{n3}$, respectively. This suggests the following trial function, as a good candidate for the variational function:

\begin{equation}
\Psi^{I_pI_n}_{M_pM_n}={\cal N}e^{zI_{p-}}e^{xI_{n-}}|I_pM_pI_p\rangle |I_nM_n I_n \rangle.
\end{equation} 
with $z$ and $x$ being complex functions of time playing  the role of classical phase space coordinates.
The term ${\cal N}$ is the normalization factor: 
\begin{equation}
{\cal N}=\left[(1+|z|^2)(1+|x|^2)\right]^{-1/2},
\end{equation}
while the states $|I_pM_pI_p\rangle$ and $|I_nM_n I_n\rangle$ are defined as follows:
\begin{eqnarray}
|I_pM_pI_p\rangle &=& \sqrt{\frac{2I_p+1}{8\pi^2}}D^{I_p*}_{M_pI_p}\nonumber\\
|I_nM_n I_n\rangle &=& \sqrt{\frac{2I_n+1}{8\pi^2}}D^{I_n*}_{M_n I_n}.
\end{eqnarray}
Throughout this paper the units where $\hbar =c =1$ are used.

Obviously, the function $\Psi^{S}_{M\sigma}$ is a linear combination of the vector states $|I_pM_pK_p\rangle ||I_nM_nK_n\rangle$ with the amplitudes close to those obtained through diagonalization procedure \cite{Rad017}. This feature appears to be determined by the fact that the variational function is a product of two coherent states for the SU2 groups generated by the components of the proton and the neutron angular momenta, respectively. Moreover, the overcompletness property of the coherent states makes possible that the spectroscopic properties of the model Hamiltonian are recovered  by re-quantizing the classical trajectories

The average value of the partial derivative has the expression:
\begin{equation}
\langle \frac{\partial}{\partial t} \rangle =I_pI_n\frac{\stackrel{\bullet}{z}z^*-z{\stackrel{\bullet}{z}}^*}{1+|z|^2}\frac{\stackrel{\bullet}{x}x^*-x{\stackrel{\bullet}{x}}^*}{1+|x|^2}
\end{equation}
where the symbol "$\bullet$" is used for the time derivative operation.

In order to get the average of H with the the trial function defined above, we need the matrix elements of the raising and lowering operators:
\begin{eqnarray}
\langle\hat{I}_{p-}\rangle &=&
{\cal N}_z^{2}\frac{\partial}{\partial{z}}({\cal N}_z^{-2})=\frac{2I_pz^{*}}{1+zz^{*}},\nonumber\\
\langle\hat{I}_{p+}\rangle &=&
{\cal N}_z^{2}\frac{\partial}{\partial{z^{*}}}({\cal N}_z^{-2})=\frac{2I_pz}{1+zz^{*}},\nonumber\\
\langle\hat{I}_{n-}\rangle &=&
{\cal N}_x^{2}\frac{\partial}{\partial{x}}({\cal N}_x^{-2})=\frac{2I_nx^{*}}{1+xx^{*}},\nonumber\\
\langle\hat{I}_{n+}\rangle &=&
{\cal N}_x^{2}\frac{\partial}{\partial{x^{*}}}({\cal N}_x^{-2})=\frac{2I_nx}{1+xx^{*}},\nonumber\\
\end{eqnarray}
The partial normalization factors have the expressions:
\begin{equation}
N_z=(1+|z|^2)^{-1/2},\;\;N_x=(1+|x|^2)^{-1/2}
\end{equation}
Also, it is easy to calculate the averages of the 3rd components of the vectors $\vec{I}_p$ and$\vec{I}_n$:
\begin{eqnarray}
\langle \hat{I}_{p3}\rangle &=& I_p-z\langle \hat{I}_{p-}\rangle =
I_p-\frac{2I_pzz^{*}}{1+zz^{*}},\nonumber\\
\langle \hat{I}_{n3}\rangle & = & I_n-x\langle \hat{I}_{n-}\rangle =
I_n-\frac{2I_nxx^{*}}{1+xx^{*}}.
\end{eqnarray}
Using the above equations, one obtains:
\begin{equation}
\langle \hat{I}_{p3}^2\rangle =\langle (\hat{I}_{p3}-z\hat{I}_{p-})(I_p-z\hat{I}_{p-})\rangle =I_p^{2}-\frac{2I_p(2I_p-1)zz^{*}}{(1+zz^{*})^{2}}.
\end{equation} 
Averages for the components  1 and 2 of the angular momentum are obtained by combining the expressions listed above:
\begin{eqnarray}
\langle\hat{I}_{p1}^{2}\rangle &=&
\frac{1}{4}\left[2I_p+\frac{2I_p(2I_p-1)}{(1+zz^{*})^{2}}(z+z^{*})^2\right],\nonumber\\
\langle\hat{I}_{2}^{2}\rangle &=&
-\frac{1}{4}\left[-2I_p+\frac{2I_p(2I_p-1)}{(1+zz^{*})^{2}}(z-z^{*})^2\right].
\end{eqnarray}
It is worth mentioning the fact that the averages of angular momenta square are $I_p(I_p+1)$ and $I_n(I_n+1)$ respectively
\begin{eqnarray}
\langle\hat{I}_{p1}^{2}\rangle+\langle\hat{I}_{p2}^{2}\rangle+\langle\hat{I}_{p3}^{2}\rangle &=& I_p(I_p+1),\nonumber\\
\langle\hat{I}_{n1}^{2}\rangle+\langle\hat{I}_{n2}^{2}\rangle+\langle\hat{I}_{n3}^{2}\rangle &=& I_n(I_n+1).
\end{eqnarray}
These equations reflect the fact that
 $\Psi(z)^{I_p,I_n}_{M_p,M_n}$ is an eigenfunction for both $\hat{I}_{p}^2$ and $\hat{I}_{n}^2$.

For what follows it is convenient to use the polar form for the classical coordinates $z$ and $x$;
\begin{equation}
z=\rho e^{i\varphi},\;\; x=\sigma e^{i\psi}.
\end{equation}
Also, we change the variables $\rho$ and $\sigma$ to:
\begin{equation}
r=\frac{2I_p}{1+\rho^2},\;\;t=\frac{2I_n}{1+\sigma^2}
\end{equation}
In terms of the new coordinates the classical energy  function is:
\begin{eqnarray}
&&\langle \Psi^{IS}_{M\sigma}|H|\Psi^{IS}_{M\sigma}\rangle \equiv {\cal H}= {\cal H}_0+{\cal H}_1 ,\;\;\rm{where} \nonumber\\
&&{\cal H}_0=A_pI_p(I_p+1)+A_nI_n(I_n+1)+V(I_p-1)(I_n-1),\nonumber\\
&&{\cal H}_1=V\left\{\left[\sqrt{rt(2I_p-r)(2I_n-t)}\cos(\psi-\varphi)\right]+\left[\frac{I_p-1}{2I_n}t+\frac{I_n-1}{2I}_pr+\frac{rt}{4I_pI_n}\right]\right\}.
\label{h0h1}
\end{eqnarray}
with the notation
\begin{equation}
A_{\tau}=\frac{1}{2{\cal J}_{\tau}}, \tau=p,n.
\end{equation}
The time dependent variational equation yields the classical equations of motion, which are of canonical Hamilton form:
\begin{eqnarray}
\frac{\partial{\cal H}}{\partial r}&=&\stackrel{\bullet}{\varphi},\;\;\frac{\partial{\cal H}}{\partial \varphi}=-\stackrel{\bullet}{r},\nonumber\\
\frac{\partial{\cal H}}{\partial t}&=&\stackrel{\bullet}{\psi},\;\;\frac{\partial{\cal H}}{\partial \psi}=-\stackrel{\bullet}{t}.
\label{Hameq}
\end{eqnarray}
Inserting the classical energy into the equations of motion (\ref{Hameq}) one obtains:

\begin{eqnarray}
&&\frac{V}{2}\left[\sqrt{\frac{t(2I_n-t)}{r(2I_p-r)}}(I_p-r)2\cos(\psi-\varphi)+ \frac{I_n-1}{I_p}+\frac{t}{2I_pI_n}\right]=\stackrel{\bullet}{\varphi},\nonumber\\
&&\frac{V}{2}\left[\sqrt{\frac{r(2I_p-r)}{t(2I_n-t)}}(I_n-t)2\cos(\psi-\varphi)+ \frac{I_p-1}{I_n}-\frac{r}{2I_pI_n}\right]=\stackrel{\bullet}{\psi},\nonumber\\
&&V\sqrt{rt(2I_p-r)(2I_n-t)}\sin(\psi-\phi)=-\stackrel{\bullet}{r},\nonumber\\
&&\;\;-V\sqrt{rt(2I_p-r)(2I_n-t)}\sin(\psi-\phi)=-\stackrel{\bullet}{t}.
\label{eqmotion1}
\end{eqnarray}
Cancelling the time derivatives, the above equations lead to a set of equations defining the stationary points for  the constant energy surface.
The stationary angles are related by:
\begin{equation}
\stackrel{\circ}{\psi}=\stackrel{\circ}{\varphi}+\pi
\end{equation}
while the stationary coordinates $\stackrel{\circ}{r}$ and $\stackrel{\circ}{t}$ are solutions of equations:
\begin{eqnarray}
&&\frac{V}{2}\left[-2\sqrt{\frac{t(2I_n-t)}{r(2I_p-r)}}(I_p-r)+ \frac{I_n}{2I_p}-\frac{1}{4I_pI_n}(2I_n-t)\right]=0,\nonumber\\
&&\frac{V}{2}\left[-2\sqrt{\frac{r(2I_p-r)}{t(2I_n-t)}}(I_n-t)+ \frac{I_p}{2I_n}-\frac{1}{4I_pI_n}(2I_p-r)\right]=0.
\label{minim}
\end{eqnarray}
It is worth mentioning that these solutions are stationary coordinates points for the constant energy function. Among them one depicts those which make the energy function minimum:
\begin{equation}
{\cal E}_{min}(r,t)=\frac{V}{2}\left[-2\sqrt{rt(2I_p-r)(2I_n-t)}+\frac{I_p-1}{2I_n}t+\frac{I_n-1}{2I_p}r+\frac{rt}{4I_pI_n}+(I_p-1)(I_n-1)\right].
\label{cale1}
\end{equation}
Note that equations of motion (\ref{eqmotion1}) are highly non-linear. However linearizing the left hand side by expanding it around the the deepest minimum and keeping
only the linear terms in the deviation, $\varphi^{\prime},\psi^{\prime}, r^{\prime}, t^{\prime}$, one obtains an integrable set of equations:
\begin{eqnarray}
\{{\cal H}, q_1\}&=&A_{11}p_1+A_{12}p_2=\stackrel{\bullet}{q}_1,\nonumber\\
\{{\cal H}, q_2\}&=&A_{12}p_1+A_{22}p_2=\stackrel{\bullet}{q}_2,\nonumber\\
\{{\cal H}, p_1\}&=&-B_{11}q_1+B_{11}q_2=\stackrel{\bullet}{p}_1,\nonumber\\
\{{\cal H}, p_2\}&=&B_{11}q_1-B_{11}q_2=\stackrel{\bullet}{p}_2.
\end{eqnarray}
Here,the current point of the phase space was denoted by;
\begin{equation}
(q_1, q_2, p_1, p_2)=(\varphi^{\prime},\psi^{\prime}, r^{\prime}, t^{\prime})
\end{equation}
while the analytical expressions of the involved coefficients, $A_{11}, A_{12}, A_{22}, B_{11}$, are given in Appendix A
For what follows it is useful to introduce the complex canonical coordinates:
\begin{eqnarray}
{\cal B}^*&=&\frac{q_1+ip_1}{\sqrt{2}},\;\;{\cal B}=\frac{q_1-ip_1}{\sqrt{2}},\nonumber\\
{\cal C}^*&=&\frac{q_2+ip_2}{\sqrt{2}},\;\;{\cal C}=\frac{q_2-ip_2}{\sqrt{2}}.
\end{eqnarray}
where $i$ denotes the imaginary unit
The complex coordinates obey the following equations of motion:
\begin{eqnarray}
\{{\cal H},{\cal B}^*\}&=&\frac{i}{2}\left[{\cal B}^*\left(-A_{11}-B_{11}\right)+{\cal C}^*\left(-A_{12}+B_{11}\right)+{\cal B}\left(A_{11}-B_{11}\right)+{\cal C}
\left(A_{12}+B_{11}\right)\right],\nonumber\\
\{{\cal H},{\cal C}^*\}&=&\frac{i}{2}\left[{\cal B}^*\left(-A_{12}+B_{11}\right)+{\cal C}^*\left(-A_{22}-B_{11}\right)+{\cal B}\left(A_{12}+B_{11}\right)+{\cal C}
\left(A_{22}-B_{11}\right)\right],\nonumber\\
\{{\cal H},{\cal B}\}&=&\frac{i}{2}\left[{\cal B}^*\left(-A_{11}+B_{11}\right)+{\cal C}^*\left(-A_{12}-B_{11}\right)+{\cal B}\left(A_{11}+B_{11}\right)+{\cal C}
\left(A_{12}-B_{11}\right)\right],\nonumber\\
\{{\cal H},{\cal C}\}&=&\frac{i}{2}\left[{\cal B}^*\left(-A_{12}-B_{11}\right)+{\cal C}^*\left(-A_{22}+B_{11}\right)+{\cal B}\left(A_{12}-B_{11}\right)+{\cal C}
\left(A_{22}+B_{11}\right)\right].
\label{eqmot2}
\end{eqnarray}
Now we map the complex coordinates onto boson operators by using the correspondence:
\begin{equation}
\left({\cal B}, {\cal B}^*,{\cal C}, {\cal C}^* \right)\to \left(b,b^{\dagger}, c, c^{\dagger}\right),\;\;\{,\}\to \frac{1}{i}\left[,\right].
\end{equation}
Indeed, according to this mapping the newly introduced  operators obey the boson-like commutation relations:
\begin{equation}
\left[b,b^{\dagger}\right]=1,\;\;\left[c,c^{\dagger}\right]=1.
\end{equation}
Moreover, the image of ${\cal H}$ through this mapping is the Hamiltonian $\hat{H}$ and Eqs. (\ref{eqmot2}) lead to the equations of motion for  the boson operators defined above. Using the resulting equations of motion for bosons, one can define the phonon operator
\begin{equation}
\Gamma^{\dagger}_{s;IS}=X_1b^{\dagger}+X_2c^{\dagger}-Y_1b-Y_2c,
\end{equation}
such that the following equations are satisfied:
\begin{equation}
\left[\hat{H},\Gamma^{\dagger}_{s;I_pI_n}\right]=\omega^{I_pI_n}_{s}\Gamma^{\dagger}_{s,I_pI_n},\;\;\left[\Gamma_{s,I_pI_n},\Gamma^{\dagger}_{s,I_pI_n}\right]=1.
\end{equation}
These restrictions imply that the phonon amplitudes are determines by the equations:
\begin{eqnarray}
&&\left(\begin{matrix}{\cal A}& {\cal B}\cr
                     -{\cal B}&-{\cal A}\end{matrix}\right)\left(\begin{matrix}X_1\cr
                                                                                X_2\cr
                                                                                Y_1\cr
                                                                                 Y_2\end{matrix}\right)
=\omega_{s}^{IS}\left(\begin{matrix}X_1\cr
                                                                                X_2\cr
                                                                                Y_1\cr
                                                                                 Y_2\end{matrix}\right),\nonumber\\
&&X_1^2+X_2^2-Y_1^2-Y_2^2=1.
\label{RPA}
\end{eqnarray}

Here ${\cal A}$ and ${\cal B}$ are $2\times 2$ matrices having the expressions:
\begin{eqnarray}
{\cal A}&=&\left(\begin{matrix}\frac{1}{2}(A_{11}+B_{11})&\frac{1}{2}(A_{12}-B_{11}) \cr
                              \frac{1}{2}(A_{12}-B_{11})&\frac{1}{2}(A_{22}+B_{11})\end{matrix}\right),\nonumber\\
{\cal B}&=&\left(\begin{matrix}\frac{1}{2}(-A_{11}+B_{11})&\frac{1}{2}(-A_{12}-B_{11}) \cr
                              \frac{1}{2}(-A_{12}-B_{11})&\frac{1}{2}(-A_{22}+B_{11})\end{matrix}\right).
\end{eqnarray}
The index "s" suggests a semi-classical treatment.
The solutions of these equations will be discussed in section 5.
\renewcommand{\theequation}{4.\arabic{equation}}
\setcounter{equation}{0}
\section{Boson description}
\subsection{The Dyson boson representation}
The components of angular momenta written in terms of the conjugate classical coordinates look like:
\begin{eqnarray}
I_{p+}&=&\sqrt{r(2I_p-r)}e^{i\varphi},\;\;I_{n+}=\sqrt{t(2I_n-t)}e^{i\psi},\nonumber\\
I_{p-}&=&\sqrt{r(2I_p-r)}e^{-i\varphi},\;\;I_{n-}=\sqrt{t(2I_n-t)}e^{-i\psi},\nonumber\\
I_{p3}&=&r-I_p,\;\;\;\;\;\;\;\;\;\;\;\;\;\;\;\;\;\;\;I_{n3}=t-I_n.
\label{I+I-}
\end{eqnarray}
Through a canonical transformation one obtains  two other pairs of canonical conjugate classical coordinates:
\begin{eqnarray}
{\cal C}_1&=&\frac{1}{\sqrt{2I_p}}\sqrt{r(2I_p-r)}e^{i\varphi};\;\;{\cal B}_1^*=\sqrt{2I_p}\sqrt{\frac{2I_p-r}{r}}e^{i\varphi},\nonumber\\
{\cal C}_2&=&\frac{1}{\sqrt{2I_n}}\sqrt{t(2I_n-t)}e^{i\psi};\;\;{\cal B}_{2}^*=\sqrt{2I_n}\sqrt{\frac{2I_n-t}{t}}e^{i\psi}.
\end{eqnarray}
Indeed, it easy to calculate the Poisson brackets of the conjugate coordinates. The result is:
\begin{equation}
\{{\cal B}_1^*,{\cal C}_1\}=i,\;\;\{{\cal B}_2^*, {\cal C}_2 \}=i,
\end{equation}
The classical complex coordinates are now quantized through the mapping:

\begin{eqnarray}
&&{\cal C}_1\to C,\;\;{\cal B}_1^*\to C^{\dagger},\;\;\{,\}\to -i[,],\nonumber\\
&&{\cal C}_2\to D,\;\;{\cal B}_2^*\to D^{\dagger},\;\;\{,\}\to -i[,].
\end{eqnarray}
Through this mapping the classical angular momenta become the angular momenta operators:

\begin{eqnarray} 
\hat{I}_{p+}&=&\sqrt{2I_p}C,\;\;\hat{I}_{p-}=\sqrt{2I_p}\left(C^{\dagger}-\frac{1}{2I_p}{C^{\dagger}}^2C\right),\;\;\hat{I}_{p3}=I_p-C^{\dagger}C,\nonumber\\
\hat{I}_{n+}&=&\sqrt{2I_n}D,\;\;\hat{I}_{n-}=\sqrt{2I_n}\left(D^{\dagger}-\frac{1}{2I_n}{D^{\dagger}}^2D\right),\;\;\hat{I}_{n3}=I_n-D^{\dagger}D.
\end{eqnarray}
In the above equations we recognize the so called Dyson boson representation for angular momenta.
Analogously, one can quantize any function defined on the classical phase space.
In particular, the classical energy (\ref{h0h1}) becomes the Hamiltonian operator:
\begin{eqnarray}
\hat{H}&=&{H}_0+\frac{V}{2}\left\{\left[2\sqrt{I_pI_n}\left(D^{\dagger}C+C^{\dagger}D\right)-\sqrt{\frac{I_n}{I_p}}{C^{\dagger}}^2CD-\sqrt{\frac{I_p}{I_n}}{D^{\dagger}}^2DC\right]+
2\left(I_p-C^{\dagger}C\right)\left(I_n-D^{\dagger}D\right)\right\},\nonumber\\
{H}_0&=&A_{p}I_p(I_p+1)+A_{n}I_n(I_n+1).
\label{Has}
\end{eqnarray}
If only the quadratic terms in bosons are retained in H, the result leads to the following equations of motion:
\begin{eqnarray}
\left[H,C^{\dagger}\right]& = &-VI_nC^{\dagger}+V\sqrt{I_pI_n}D^{\dagger},\nonumber\\
\left[H,D^{\dagger}\right]& = &V\sqrt{I_pI_n}C^{\dagger}-VI_pD^{\dagger},\nonumber\\
\left[H,C\right]&=&VI_nC-V\sqrt{I_pI_n}D,\nonumber\\
\left[H,D\right]&=&-V\sqrt{I_pI_n}C+VI_pD.
\label{eqmot}
\end{eqnarray}
Further we define the phonon operator:
\begin{equation}
\Gamma^{\dagger}=X_1C^{\dagger}+X_2D^{\dagger}-Y_1C-Y_2D,
\end{equation}
such that the following restrictions are obeyed:
\begin{equation}
\left[H,\Gamma^{\dagger}\right]=\omega \Gamma^{\dagger},\;\;\left[\Gamma,\Gamma^{\dagger}\right]=1.
\label{eqga}
\end{equation}
The first equation, from above, yields a homogeneous system of equations for phonon  amplitudes, whose compatibility restriction determines the energy $\omega$:
\begin{equation}
\omega^2\pm V(I_p+I_n)\omega =0.
\end{equation}
There are two non-vanishing solutions for $\omega$:
\begin{equation}
\omega=\mp V(I_p+I_n).
\end{equation}
Note that for an attractive interaction  the classical energy exhibits a minimum value  if  $\vec{I}_p$ is parallel with $\vec{I}_n$ which actually is expected to happen for high angular momenta. In this case the first solution is valid.  By contrary, for a repulsive interaction the anti-alignment of the proton and neutron angular momenta
is favored in the low spin region.  In this case the second solution is acceptable. However in such a case the phonon operator cannot be normalized to unity, i.e. the second equation (\ref{eqga}) is not obeyed. Concluding none of above solutions for $\omega$ is acceptable.

The approximation described above can be improved by involving some contribution coming from the quartic boson terms of $H$, through the Bogoliubov transformation:
\begin{equation}
C^{\dagger}=U_1\tilde{C}^{\dagger}-V_1\tilde{C},\;\;D^{\dagger}=U_2\tilde{D}^{\dagger}-V_2\tilde{D},
\end{equation}
where the new operators are bosons if the coefficients $U$ and $V$ obey the normalization conditions:
\begin{equation}
U_1^2-V_1^2=1,\;\;U_2^2-V_2^2=1.
\end{equation} 
Witting the cubic operators in terms of the new bosons and then performing  a normal ordering of the result and keeping only the linear terms one gets:
\begin{eqnarray}
{C^{\dagger}}^2C& = &3U_1V_1^2\tilde{C}^{\dagger}-\left(U_1^2V_1+2V_1^3\right)\tilde{C},\nonumber\\
{D^{\dagger}}^2D& = &3U_2V_2^2\tilde{D}^{\dagger}-\left(U_2^2V_2+2V_2^3\right)\tilde{D}.
\label{C3DD3C}
\end{eqnarray}
The independent coefficients , say, $V_1$ and $V_2$ are fixed such that the cross terms $\tilde {C}^{\dagger}\tilde {C}^{\dagger}+\tilde {C}\tilde {C}$
and $\tilde {D}^{\dagger}\tilde {D}^{\dagger}+\tilde {D}\tilde {D}$ are cancelled. These restrictions lead to:
\begin{equation}
V_1^2=I_p,\;\;V_2^2=I_n.
\label{V1V2}
\end{equation}
Inserting  (\ref{C3DD3C}) into  (\ref{Has}) one obtains a Hamiltonian that is quadratic in the new bosons. The salient feature of Dyson boson expansion is that it is a finite
expansion. However, the drawback is  that it does not preserve hermiticity,i.e. a hermitian operator becomes, after expansion, non-hermitian. This is the case of our Hamiltonian. Moreover, this picture is still valid even after the linearization. Despite the fact that the transformed Hamiltonian is non-hermitian, it has real eigenvalues \cite{Ogu}. Since it is not comfortable at all to diagonalize a non-Hermitian operator we prefer instead to use, from this point onward, the Hermitian operator:
\begin{equation}
\bar{H}=\frac{1}{2}(\hat{H}+\hat{H}^{\dagger}),
\end{equation}
which admits the same eigenvalues as H.
The equations of motion for the tilde operators are:
\begin{eqnarray}
&&\left[\bar{H},\tilde{C}^{\dagger}\right]=a\tilde{D}^{\dagger}+b\tilde{D},\nonumber\\
&&\left[\bar{H},\tilde{D}^{\dagger}\right]=a\tilde{C}^{\dagger}+b\tilde{C},\nonumber\\
&&\left[\bar{H},\tilde{C}\right]=-b\tilde{D}^{\dagger}-a\tilde{D},\nonumber\\
&&\left[\bar{H},\tilde{D}\right]=-b\tilde{C}^{\dagger}-a\tilde{C}.
\end{eqnarray}
where the following notations have been used:
\begin{eqnarray}
a&=&-\sqrt{I_p(I_p+1)I_n(I_n+1)}-I_pI_n -\frac{1}{2}(I_p+I_n),\nonumber\\
b&=&\frac{1}{2}\sqrt{I_p(I_p+1)}(2I_n+1)+\frac{1}{2}\sqrt{I_n(I_n+1)}(2I_p+1).
\end{eqnarray}
The phonon operator
\begin{equation}
\Gamma^{\dagger}_{b;I_pI_n}=X_1\tilde{C}^{\dagger}+X_2\tilde{D}^{\dagger}-Y_1\tilde{C}-Y_2\tilde{D}.
\end{equation}
is determined such that the following equations are fulfilled:
\begin{eqnarray}
\left[\bar{H},\Gamma^{\dagger}_{b;I_pI_n}\right]&=&\omega_b^{I_pI_n}\Gamma^{\dagger}_{b,I_pI_n},\nonumber\\
\left[\Gamma_{b,I_pI_n},\Gamma^{\dagger}_{b;I_pI_n}\right]&=&1.
\end{eqnarray}
There exists only one positive solution of the above equations:
\begin{eqnarray}
&&\omega_{b}^{I_pI_n}=|a|\frac{V}{2},\nonumber\\
&&X_1=\left(2-\frac{b^2}{a^2}\right)^{-1/2},\;\;X_2=-X_1,\nonumber\\
&&Y_1=-\frac{b}{a}X_1,\;\;Y_2=0.
\end{eqnarray}

\renewcommand{\theequation}{5.\arabic{equation}}
\setcounter{equation}{0}
\section{Magnetic dipole transitions}
The spherical components of magnetic dipole transition operator have the expressions:
\begin{equation}
M_{1\mu}=\sqrt{\frac{3}{4\pi}}(g_pI_{p\mu}+g_nI_{n\mu}),
\end{equation}
where $g_p, g_n$ denote the protons and neutrons gyromagnetic factor, respectively.

\subsection{Semi-classical description}
The states $| I^+\rangle$ forming the band (I,1) are vacuum states for the phonon operator (3.27) corresponding to the energy $\omega_s^{I,1}$. Due to this remark we use the notation $|0\rangle_{s,I}$ for the state $| I^+\rangle$. One can easily check that the phonon operator is a tensor of rank 1.

We are interested in calculating the transition between the states $|o\rangle_{s,I}$ and $|(1+1)^+\rangle = \Gamma^{\dagger}_{1,\mu}|0\rangle_{s,I} $.
One can easily check that  in order to calculate the matrix elements characterizing the transition $|1^+\rangle \to |(I+1)^+\rangle$ we have first to expand, in the first order, the angular momenta components  (\ref{I+I-}) around  the energy function minimum and then quantize the deviations according to the procedure described in Section 3. Results for the first phonon state decay are as follows:
\begin{eqnarray}
{}_{s,I_p}\langle 0|{I_p}_{+1}\Gamma^{\dagger}_{1,-1}|0\rangle_{s,I_p} &=&\frac{-i}{2}\left[\frac{I_p-\stackrel{\circ}{r}}{\sqrt{\stackrel{\circ}{r}(2I_p-\stackrel{\circ}{r})}}\left(Y_{1}-X_{1}\right)
-\sqrt{\stackrel{\circ}{r}(2I_p-\stackrel{\circ}{r})}\left(Y_{1}+X_{1}\right)\right],\nonumber\\
{}_{s,I_p}\langle 0|{I_{p-}}_{-1}\Gamma^{\dagger}_{1,1}|0 \rangle_{s,I_p} &=&\frac{-i}{2}\left[\frac{I_p-\stackrel{\circ}{r}}{\sqrt{\stackrel{\circ}{r}(2I_p-\stackrel{\circ}{r})}}\left(Y_{1}-X_{1}\right)
+\sqrt{\stackrel{\circ}{r}(2I_p-\stackrel{\circ}{r})}\left(Y_{11}+X_{11}\right)\right],\nonumber\\
{}_{s,I_p}\langle 0|I_{p0}\Gamma^{\dagger}_{1,0}|0\rangle_{s,I_p} &=&\frac{i}{\sqrt{2}}\left(X_{1}-Y_{1}\right),\nonumber\\
{}_{s,I_p}\langle 0|{I_n}_{+1}\Gamma^{\dagger}_{1,-1}|0\rangle_{s,I_p} &=&\frac{-i}{2}\left[\frac{I_n-\stackrel{\circ}{t}}{\sqrt{\stackrel{\circ}{t}(2I_n-\stackrel{\circ}{t})}}\left(Y_{2}-X_{2}\right)
-\sqrt{\stackrel{\circ}{t}(2I_n-\stackrel{\circ}{t})}\left(Y_{2}+X_{2}\right)\right],\nonumber\\
{}_{s,I_p}\langle 0|{I_n}_{-1}\Gamma^{\dagger}_{1,1}|0\rangle_{s,I_p}&=&\frac{-i}{2}\left[\frac{I_n-\stackrel{\circ}{t}}{\sqrt{\stackrel{\circ}{t}(2I_n-\stackrel{\circ}{t})}}\left(Y_{21}-X_{21}\right)
+\sqrt{\stackrel{\circ}{t}(2I_n-\stackrel{\circ}{t})}\left(Y_{2}+X_{2}\right)\right],\nonumber\\
{}_{s,I_p}\langle 0|{I_n}_{0}\Gamma^{\dagger}_{1,0}|0\rangle-[s,I_p] &=& \frac{i}{\sqrt{2}}\left(X_{2}-Y_{2}\right).
\end{eqnarray}
Using these matrix elements, the reduced probability for  the M1 dipole transition  from a I state of the ground band to the $(I+1)^+$ state of the excited band,is readily obtained
\begin{equation}
B(M1;I_p^+\to (I_p+1)^+)=\sum_{\mu}|{}_{s,I_p}\langle 0|\Gamma_{s;I_p1,\mu} M_{1}\mu|0\rangle_{s,I_p}|^2.
\end{equation}

\subsection{The full Dyson boson expansion}
The linearized boson expansions associated to the a.m. components are:
\begin{eqnarray}
{I^D_p}_+&=&\sqrt{2I_p}\left(U_1\tilde{C}-V_1\tilde{C}^{\dagger}\right),\nonumber\\
{I^D_p}_-&=&\sqrt{2I_p}\left[\left(U_1-\frac{3}{2I_p}U_1V_1^2\right)\tilde{C}^{\dagger}+\left(-V_1++\frac{1}{2I_p}(U_1^2V_1+2V_1^3)\right)\tilde{C}\right],\nonumber\\ 
{I^D_n}_+&=&\sqrt{2I_n}\left(U_2\tilde{D}-V_2\tilde{D}^{\dagger}\right),\\
{I^D_n}_-&=&\sqrt{2I_n}\left[\left(U_2-\frac{3}{2I_n}U_2V_2^2\right)\tilde{D}^{\dagger}+\left(-V_2++\frac{1}{2I_n}(U_2V_2+2V_2^3)\right)\tilde{D}\right].\nonumber 
\end{eqnarray}
We note that the hermiticity property is broken,i.e. $({I^D_p}_+)^{\dagger}\neq {I^D_p}_-$ and $({I^D_n}_+)^{\dagger}\neq {I^D_n}_-$.

To calculate the M1 transition probability for the case when the full Dyson boson expansion is used for angular momenta, we need the following matrix element:
\begin{eqnarray}
{}_{b,I_p}\langle 0|{I^D_p}_{+1}\Gamma^{\dagger}_{1,-1}|0\rangle_{b,I_p} &=&-\sqrt{I_p}\left(U_1X_{1}-V_1Y_{1}\right),\nonumber\\
{}_{b,I_p}\langle 0|I^D_{-1}\Gamma^{\dagger}_{1,+1}|0\rangle_{b,I_p} &=&\sqrt{I_p}\left[-V_1X_{1}+U_1Y_{1}-\frac{1}{2I_p}\left(3U_1V_1^2Y_{1}-(U_1^2V_1+2V_1^3)X_{1}\right)\right],\nonumber\\
{}_{b,_p}\langle 0|{I^D_n}_{+1}\Gamma^{\dagger}_{1,-1}|0\rangle_{b,I_p} &=&-\sqrt{I_n}\left(U_2X_{2}-V_2Y_{2}\right),\nonumber\\
{}_{b,I_p}\langle 0|{I^D_n}_{-1}\Gamma^{\dagger}_{1,+1}|0\rangle_{b,I_p} &=&\sqrt{I_n}\left[-V_2X_{2}+U_2Y_{2}-\frac{1}{2I_n}\left(3U_2V_2^2Y_{2}-(U_2^2V_2+2V_2^3)X_{2}\right)\right].
\end{eqnarray}
Here, the notations ${I^D_p}_{\pm 1}$ and ${I^D_n}_{\pm 1}$  are used for the spherical components of the angular momenta $\vec{I}_p$ and $\vec{I}_n$, respectively.

Using these matrix elements the reduced transition probability is readily obtained 
\begin{equation}
B(M1;I_p^+\to (I_p+1)^+)=\sum_{\mu}|{}_{b,I_p}\langle 0|\Gamma_{b;I_p1,\mu} M_{1}\mu|0\rangle_{b,I_p}|^2.
\end{equation}

\renewcommand{\theequation}{6.\arabic{equation}}
\setcounter{equation}{0}
\section{Numerical analysis and discussion}
The model Hamiltonian involves three parameters; these are the moments of inertia for protons and neutrons respectively and the rotor's interaction V'. 
 The experimental value for the moment of inertia of nuclei can be approximated by the compact formula \cite{Ring}
\begin{equation}
{\cal J}_{exp} =\frac{\beta^2A^{7/3}}{400}[\hbar^2MeV^{-1}].
\label{inert}
\end{equation}
where $\beta$ denotes the nuclear quadrupoole deformation and $A$ is the nuclear mass number.
 The result is multiplied successively by the factors $Z/A$ and $N/A$ as to obtain the proton and neutron moment of inertia,
respectively.The obvious notations for nuclear charge ($Z$) and neutron number ($N$) are used. Application concerns the isotope $^{156}$Gd for which $\beta$=0.266 
\cite{LRla} and $A$=156. The interaction strength is taken such that the  difference between the calculated energies  of the states $2^+$ from the yrast and the first excited bands respectively, be equal to 0.1 MeV. Thus one obtains:

$${\cal J}_p=9.3075  \rm{\hbar^2MeV^{-1}},\;\; {\cal J}_n=13.6649  \rm{\hbar}^2\rm{MeV}^{-1},\;\;V=0.186 \rm{MeV}$$.

The stationary values for the coordinates r,t, are obtained by solving the equations (\ref{minim}). Since the corresponding Hessian is positive the solutions define the minimum for the energy function. Results are collected in Table 1 for two sets of pairs $(I_p,I_n)$, namely $(I_p,1)$ with $I_p=1,2,3,..,10$ and $(1,I_n)$ with $I_n=1,2,3,...,10$.
The minima of the energy ${\cal E}_{min}$ (3.20), as a function of the coordinates $r$ and $t$, are depicted in Fig. 1 though a contour plot for four representative pairs of angular momenta, $(I_p,I_n)$. Note a certain symmetry in Table 1. Indeed, $(\stackrel{\circ}{r},\stackrel{\circ}{t})$ for the band ($I_p$,1) are equal with
$(\stackrel{\circ}{t},\stackrel{\circ}{r})$ for (1,$I_n$). This degeneracy reflects the invariance of the energy function tor the interchange of $I_p$ with $I_n$ and $r$ with $t$,i.e the proton neutron permutation.  

As for the stationary angles they satisfy the relation $\stackrel{\circ}{\psi}-\stackrel{\circ}{\varphi}=\pi$. We made the option for
$\stackrel{\circ}{\psi}=\pi$ and $\stackrel{\circ}{\varphi}=0$.

These data are further used to calculate the wobbling frequency, by solving the RPA-like equation (\ref{RPA}) ,for the semi-classical framework and (4.21) for the
boson description.
Wobbling frequencies, thus obtained, are employed to calculate the energies in the ground band and the one-phonon excited band:
\begin{eqnarray}
E_{s,1}^{I_p1}&=&\left[A_pI_p(I_p+1)+A_nI_n(I_n+1)\right]+{\cal E}_[min](\stackrel{\circ}{r},\stackrel{\circ}{t})
                                +\frac{1}{2}\omega_s^{I_p1},\nonumber\\
E_{s,2}^{I_p+1,1}&=&\left[A_pI_p(I_p+1)+A_nI_n(I_n+1)\right]+{\cal E}_{min}(\stackrel{\circ}{r},\stackrel{\circ}{t})+\frac{3}{2}\omega_s^{I_p1},\nonumber\\
E_{b,1}^{I_p1}&=&\left[A_pI_p(I_p+1)+A_nI_n(I_n+1)\right]+\frac{1}{2}\omega_b^{I_p1},\nonumber\\
E_{b,2}^{I_p+1,1}&=&\left[A_pI_p(I_p+1)+A_nI_n(I_n+1)\right]+\frac{3}{2}\omega_b^{I_p1},
\end{eqnarray}
where ${\cal E}_1(\stackrel{\circ}{r},\stackrel{\circ}{t})$ denotes the energy function defined by Eq. (3.20) considered in the minimum point, while $A_{\tau}$, $\tau=p,n$ is half the reciprocal proton, neutron moment of inertia, respectively.
\begin{figure}[ht!]
\includegraphics[width=0.4\textwidth]{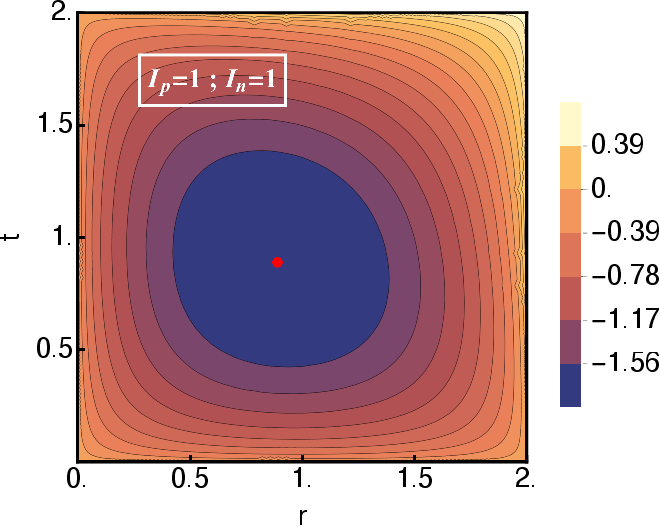}\includegraphics[width=0.4\textwidth]{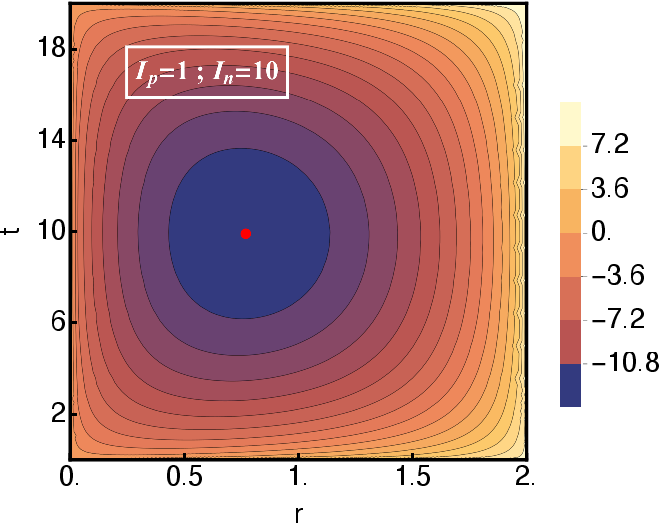}
\includegraphics[width=0.4\textwidth]{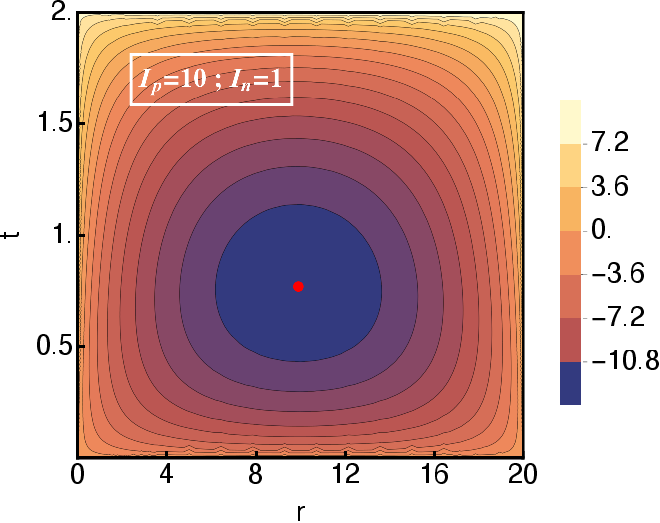}\includegraphics[width=0.4\textwidth]{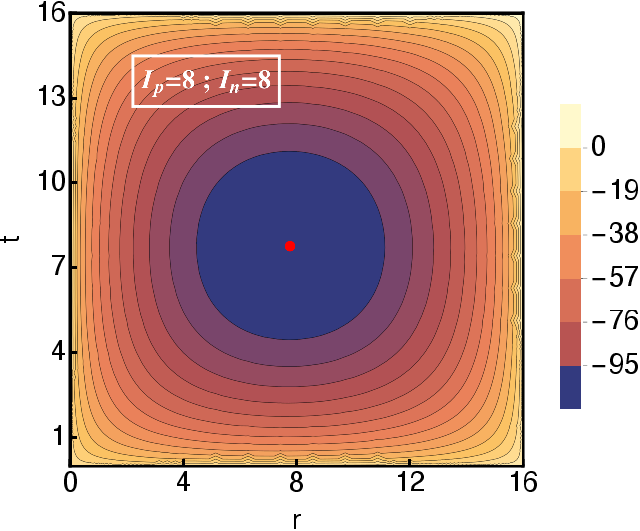}
\caption{Color online. Contour plots for four angular momenta pairs, $(I_p,I_n)$, and the energy function (3.20)}
\end{figure}
\begin{table}[hb!]
\begin{tabular}{|c|ccccc|}
\hline
$(I_p,I_n)$&$\stackrel{\circ}{r}$ &$\stackrel{\circ}{t}$&$(1,I_n)$& $\stackrel{\circ}{r}$ &$\stackrel{\circ}{t}$\\
\hline
(1,1)&      0.888889&0.888889&(1,1)&0.888889&0.888889\\
(2,1)&      1.89609&0.81856&(1,2)&0.81856&1.89609\\
(3,1)&      2.89829&0.79726&(1,3)&0.79726&2.89829\\
(4,1)&      3.89935&0.786972&(1,4)&0.786972&3.89935\\
(5,1)&      4.89998&0.780911&(1,5)&0.780911&4.89998\\
(6,1)&      5.90039&0.776915&(1,6)&776915&5.90039\\
(7,1)&      6.90068&0.774083&(1,7)&0.774083&6.90068\\
(8,1)&      7.9009&0.771971&(1,8)&0.771971&.9009\\
()9,1) &      8.90107&0.770335&(1,9)&0.770335&8.90107\\
(10,1)&     9.9012&0.769031&(1,10)&0.769031&9.9012\\
\hline
\end{tabular}
\caption{The coordinates r and t for the minimum point of the energy surface defined by Eq. 3.20..}
\end{table}

The wobbling frequencies and the energies of the ground and one-phonon excited bands as obtained from the semi-classical and boson descriptions respectively, are presented in Table II and illustrated in Fig, 2. We note that predictions of the two formalisms are close to each other.

\begin{table}[h!]
\begin{tabular}{|c|cccccc|}
\hline
$(I_p,I_n)$&$\omega_{s}^{I_p1}$&$\omega_{b}^{I_p1}$&$E_{s1}^{I_p1}$&$E_{b1}^{I_p1}$&$E_{s2}^{I_p1}$&$E_{b2}^{I_p1}$\\
     &[MeV]    & [MeV]    & [MeV]  &  [MeV]& [MeV]  & [MeV]\\
\hline
(1,1)&0.248&0.372 &0.137 &0.295 &  -  &     -    \\
(2,1)&0.411& 0.643& 0.285& 0.643&0.385&0.661 \\
(3,1)&0.592&0.924&0.546 &1.095&0.697 & 1.291\\
(4,1)&0.778&1.193& 0.915& 1.652&1.138 &2.016\\
(5,1)&0.967&1.465&1.390&2.314&1.693 &2.846 \\
(6,1)&1.156&1.736& 1.971&3.081&2.357&3.778\\
(7,1)&1.346&2.008&2.657&3.927&3.127& 4.817\\
(8,1)&0.846&2.279& 3.917& 4.930&4.003 &5.960 \\
(9,1)&1.728&2.250& 4.345&6.012 &4.766&7.209  \\
(10,1)&1.919&2.822&5.348&7.199&6.073&8.564\\
\hline
\end{tabular}
\caption{Wobbling frequencies and energies for the first (ground) and second (one-phonon excited) band  are given in units of MeV for I running from 1 to 10.}
\end{table}

\begin{figure}[ht!]
\includegraphics[height=5.5cm,width=0.3\textwidth]{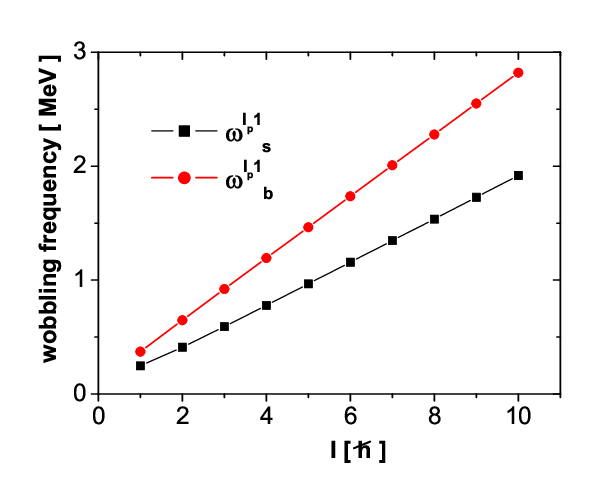}\includegraphics[height=5.5cm,width=0.3\textwidth]{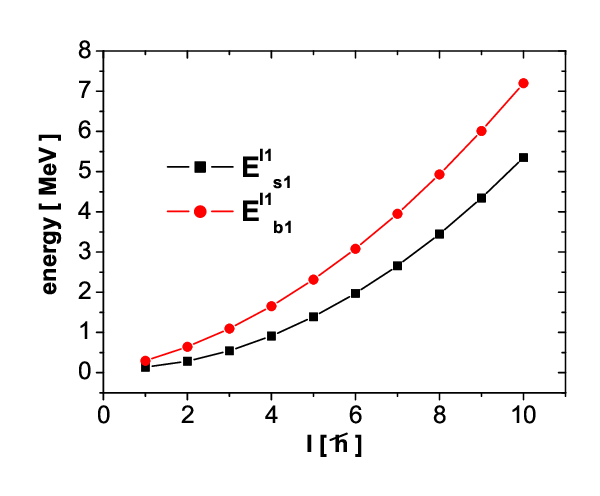}\includegraphics[height=5.5cm,width=0.3\textwidth]{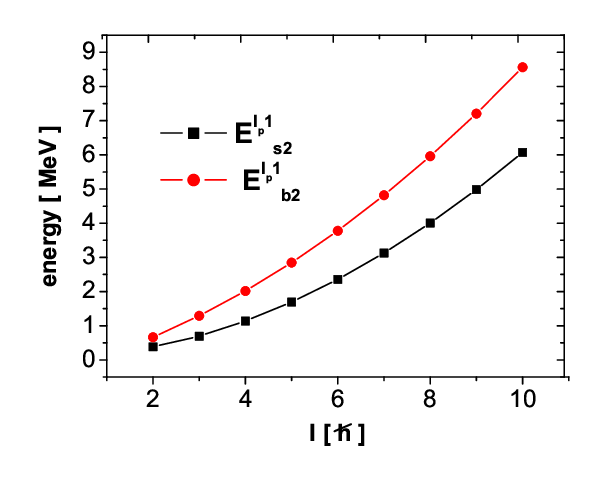}
\caption{Color online. Semi-classical and boson wobbling frequencies($\omega_s^{I_p1}$ and $\omega_b^{I_p1}$), the first band energies( $E_{s,1}^{I_p1}$  and $E_{b,1}^{I_p1}$)  and energies in the  one phonon excited band ($E_{s2}^{I_p1}$ and $E_{b2}^{I_p1}$)}.
\end{figure}

The amplitudes of the phonon operator are used now to calculate the reduced probability to perform the $M1$ transition $I^+\to (I+1)^+$.. Aiming at this goal the raising and lowering operators $I_{p\pm}$ and $I_{n\pm}$ are expressed as linear combination of the boson operators composing the dipole phonon operator.Results are given in 
Table III. Calculations were performed taking for the total gyromagnetic factor the value of $Z/A$ and $g_n=-0.1g_p$ From Table III it can be seen that the B(M1) values are increasing with respect to I. Comparing the predictions of the two formalisms one remarks that boson expansion results are
by a factor of $2\div 30$ larger than those corresponding to the semi-classical treatment. The reason is that the linearization procedures for the transition operators
used in the two cases are different. It seems that Bogoliubov transformation affects the phonon amplitudes in a more efficient way.

\begin{table}[h!]
\begin{tabular}{|c|c c|}
\hline 
I &\multicolumn{2}{c|}{$B(M1;I^+\to(I+1)^+$)[$\mu_N^2$]} \\
\hline
 &semi-classic& Dyson boson \\
\hline
1&0.0291 &  0.0270 \\  
2&0.0280 &   0.0424\\
3&0.0279& 0.0718   \\
4&0.0283 & 0.1215    \\
5&0.0259 &0.1901   \\
6&0.0249 &0.3083   \\
7&0.0301 &0.4505     \\
8&0.1146 & 0.6556   \\
9&0.0313 &0.9060 \\
\hline
\end{tabular}
\caption{The B(M1) values for the dipole transitions $I_p^+\to (I_p+1)^+$ are given for $I_p$=1,2,3,..,9, in units of $\mu_N^2$.}
\end{table}

A natural question arises about the nature of these states. Are they scissors or shears modes? To answer this question one has to calculate the angle between 
$\vec{I}_p$ and $\vec{I}_n$. The value for  this angle is obtained by solving the equation:
\begin{equation}
\cos(\theta_{I_pI_n})=\frac{\vec{I}_p\cdot\vec{I}_n}{|\vec{I}_p||\vec{I}_n|}=\frac{\langle I\rangle(\langle I\rangle+1)-I_p(I_p+1)-I_n(I_n+1)}{\sqrt{I_p(I_p+1)I_n(I_n+1)}},
\end{equation}

As for the angle between the angular momenta $\angle(\vec{I_p},\vec{I}$ and $\angle(\vec{I_n},\vec{I}$. these are  solutions of equations:
\begin{eqnarray}
\cos(\theta_{I_pI})&=&\frac{\langle I\rangle(\langle I\rangle+1)+I_p(I_p+1)-I_n(I_n+1)}{\sqrt{I_p(I_p+1)\langle I\rangle(\langle I\rangle+1)}},\nonumber\\
\cos(\theta_{I_nI})&=&\frac{\langle I\rangle (\langle I\rangle+1)+I_n(I_n+1)-I_p(I_p+1)}{\sqrt{I_n(I_n+1)\langle I\rangle (\langle I\rangle+1)}},
\end{eqnarray}

Concerning the mean value of the total angular momentum, $\langle I ]\rangle$, this obeys the following equation:
\begin{eqnarray}
&&\langle I\rangle(\langle I\rangle+1)=I_p(I_p+1)+I_n(I_n+1)+F(I_p,I_n,\stackrel{\circ}{r}\stackrel{\circ}{t}),\;\;\rm{where}\nonumber\\
&&F(I,S,\stackrel{\circ}{r},\stackrel{\circ}{t})=\\
&&\left[-2\sqrt{\stackrel{\circ}{r}\stackrel{\circ}{t}(2I_p-1)(2I_n-1)}+(I_p-1)(I_n-1)]+\frac{I_p-1}{2I_n}\stackrel{\circ}{t}
+\frac{I_n-1}{2I_p}\stackrel{\circ}{r}+\frac{\stackrel{\circ}{r}\stackrel{\circ}{r}}{4I_pI_n}\right]
\end{eqnarray}

Results of this analysis are shown in Table IV. From there we see that the angle between $\vec{I_p}$ and $\vec{I_n}$ vary slowly around $120^{\circ}$ while that between $\vec{I}_p$ and $\vec{I}$ is decreasing from  $58^{\circ}$ to about $7^{\circ}$. One notices that for the states ($I_p$,1) have a shears character. This is different from the prediction of two rotor model where the state 
$1^+$ has a scissors character.  On the last row of Table IV we listed the results of $I_p=8$ and $I_n=8$ This state is also of shears type, as any other state with $I_p=I_n$. Otherwise the other features mentioned for the $(I_p,1)$ states are still valid. Since the rotational energy is large in this case, the wobbling frequencies and the band energies are large as well. This aspect  is outlined in the following table:
\begin{eqnarray}
&&\omega_s^{88}=1.651,\rm{MeV}\nonumber\\
&&E_{s,1}^{88}=8.676\rm{MeV},\;\;E_{s,2}^{99}=10.327\rm{MeV},\nonumber\\
&&B(M1;8^+\to 9^+)=0.503 \mu_N^2,\nonumber\\
&&\omega_b^{88}=7.796,\rm{MeV}\\
&&E_{b,1}^{88}=7.823\rm{MeV},\;\;E_{b,2}^{91}=15.619\rm{MeV},\nonumber\\
&&B(M1;8^+\to 9^+)=0.1134\mu_N^2.\nonumber
\end{eqnarray}
These results correspond to V=0.108428 MeV.
\begin{table}[h!]
\begin{tabular}{|c|cccc|}
\hline
$(I_p,I_n)$ &   $\angle(\vec{I_p},\vec{I_n})$ & $\angle(\vec{I}_p,\vec{I})$ &$ \angle(\vec{I}_n,\vec{I}) $& $\langle I\rangle$\\
      &        [$^{\circ}]$             &      [$^{\circ}]$                &      [$^{\circ}]$            &       [$\hbar$]\\
\hline  
(1,1)&116.388& 58.194&58.194&1.072 \\
(2,1)&118.780&35.017&83.663&1.719 \\
(3,1)&119.887&23.951 &95.929&2.561 \\
(4,1)&120.615 & 17.973 &lined 102.462 & 3.476\\
(5,1)&121.099 & 14.316 & 106.788& 4.464\\
(6,1)&121.445 & 11.864 &109.581&5.381\\
(7,1)&121.705& 10.121& 111.584& 6.365 \\ 
(8,1)&99.457&9.594&89.863&7.885\\
(9,1)&122.067&7.8111&  122.067& 8.832\\ 
(10,1)&122.199&7.008&115.199&9.327\\
(8,8) & 119.936&59.968&59.968&8.008\\
\hline
\end{tabular}
\caption{Angles for the pairs of vectors ($\vec{I_p}.\vec{I_n})$  ($\vec{I}_p.\vec{I}$) and ($\vec{I}_n.\vec{I}$) with I denoting the total angular momentum, the average proton and neutron angular momenta corresponding to a state of angular momenta $I$ and $S=1$, respectively.}
\end{table}

Obviously, results for energy strongly depend on the proton and neutron moment of inertia. To stress on this aspect we give an example where the nuclear moment of inertia is chosen such that the energy  of the K=2 state $2^+$, i.e. the head state of the band $\gamma$ \cite{Fin}, is reproduced. Thus, one obtains 
${\cal J}_p=0.35306 \hbar^2MeV^{-1}$,
${\cal J}_n=0.50756 \hbar^2MeV^{-1}$ and $V=0.3528923 $MeV. Results for wobbling energies and the first energy levels of the ground and excited band are synthesized in the following array: 
\begin{eqnarray}
&&\omega_s^{11}=0.4702,\rm{MeV}\nonumber\\
&& E_{s,1}^{11}=4.72416\rm{MeV},\;\;E_{s,2}^{21}=5.19468\rm{MeV},\\
&{}& B(M1;1^{+} \to 2^{+})=0.029 \mu_N^2,\nonumber\\
&& \omega_b^{11}=0.70586 \rm{MeV},\nonumber\\
&& E_{b,1}^{11}=5.97847\rm{MeV},\;\;E_{b,2}^{21}=6.68427\rm{MeV},\nonumber\\
&& B(M1;1^+\to 2^+)=0.027 \mu_N^2.
\end{eqnarray}
It is  worth noting that energy for  the state $(I_p=1,I_n=1)$ obtained within the semi-classical approach, is not far from the energy of the scissors mode, which is equal to 3.075 MeV \cite{BRS83}. 

As we already mentioned there exits  a band (1,$I_n$) which is degenerate with the band ($I_p$,1). The degeneracy is caused by the invariance of the Hamiltonian to the interchange of $I_p$ with $I_n$ and  $r$ with $t$. Moreover, there exists another symmetry which is broken, namely the chiral one. The degeneracy induced by the  invariant    Hamiltonian $H_0$ is lifted up by adding the V term which breaks the  chiral symmetry. 

It can be checked that this induces an energy split
\begin{equation}
\Delta E_I=\frac{V}{2}(\langle I \rangle (\langle I \rangle)+1)-I_p(I_p+1)-I_n(I_n+1)).
\end{equation}
For the strength V=0.186 MeV, the above split ranges from 0.165 MeV for $I_p$=2 to 1.47 MeV for $I_p$=10. The resulting non-degenerate bands  exhibit the features of a chiral twin doublet.
\begin{figure}[h!]
\includegraphics[height=5.5cm,width=0.5\textwidth]{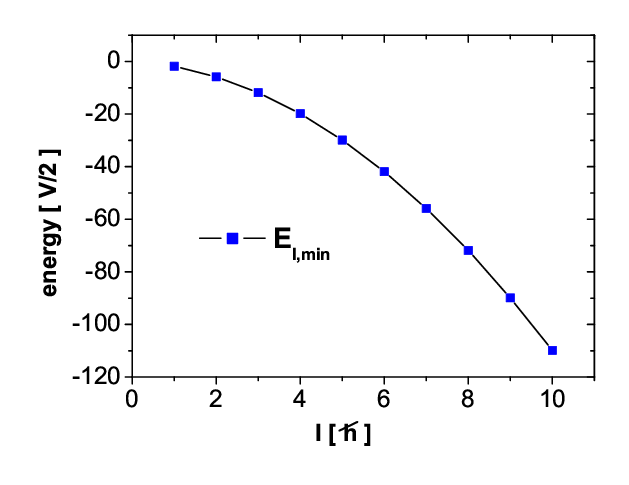}
\caption{Color online. The minimum value for the semi-classical energy function given by Eq.(3.20), considered for $I_p=I_n=I$, is plotted in units of [V/2].} 
\label{eminim}
\end{figure}

Coming back to Table IV we notice that for ($I_p,I_n$)=(1,1) and (8,8) respectively, the mean value of the total angular momentum is almost identical with $I_p$. This suggests that for $I_p=I_n$, $I_p$ approximates very well the total angular momentum. The question is whether this is generally true. This question is positively answered by our investigation shown in Table V. Indeed, the values of $\langle I \rangle $ shown in the second last column approximate very well the common values of $I_p$
 and $I_n$. Moreover it can be easily checked that the dispersion for the total angular momentum I is almost equal to zero and consequently I can be considered as a good quantum  number which, therefore, legitimate the use of label I for the state (I,I). Since in this scenario $\angle(\vec{I}_p,\vec{I}_n)\approx 120^0$ we may say that the states have a shears character.

 A specific feature of this approach is that the total angular momentum is oriented along the bisector line of the angle $\angle (\vec{I}_p,\vec{I}_n)$.

Changing the orientation either of $I_p$ or of $\vec{I}_n$ one obtains the chiral partner band of the non-excited band. As shown in Table V, the energy distance between the two partner bands is 
an increasing function of $I$, ranging from 0.089 MeV  for $I=1$ to 5.531 MeV for $I=10$.
\begin{table}[ht!]
\begin{tabular}{|c|cccccc|}
\hline
$(I_p,I_n)$ & $\stackrel{\circ}{r}=\stackrel{\circ}{t}$&  $\angle(\vec{I_p},\vec{I_n})$ & $\angle(\vec{I}_p,\vec{I})$ &$ \angle(\vec{I}_n,\vec{I}) $& $\langle I\rangle$&
$\Delta$ E\\
      &   &     [$^{\circ}]$             &      [$^{\circ}]$                &      [$^{\circ}]$            &       [$\hbar$]& [MeV]\\
\hline  
(1,1)&0.8889& 116.387 &58.194 & 59.134 &1.091 &0.089\\
(2,2)&1.8182& 119.003 &59.501 &59.501 &2.036 &0.293\\
(3,3)&2.7945&119.547&59.773&59.773&3.023&0.596\\
(4,4)&3.7829&119.743&59.872 &59.872&4.017&0.999\\
(5,5)&4.7761&119.836&59.918&59.918&5.014&1.503\\
(6,6)&5.7716&19.886&59.942&59.942&6.011&2.107\\
(7,7)&6.7684 &119.916 &59.958 &59.958 &7.009 &2.812\\
(8,8)&7.7661&119.936&59.968&59.968&8.008&3.618\\
(9,9)& 8.7642 & 119.949 & 59.968 & 59.968 & 9.007 & 4.523\\
(10,10)&9.7628 &119.956 &59.974 &59.974 &10.006 &5.531\\
\hline
\end{tabular}
\caption{The angular momenta for protons and neutrons respectively, the minimum point $(\stackrel{\circ}{r},\stackrel{\circ}{t} )$,the angle for the pairs of vectors
$(\vec{I}_p,\vec{I}_n), (\vec{I}_p,\vec{I}_n),(\vec{I}_p,\vec{I}_n)$, the mean value for the total angular momentum $\langle I \rangle $ and  the energy split for the twin chiral bands.}
\end{table}

\begin{figure}[ht!]
\includegraphics[height=5.5cm,width=0.5\textwidth]{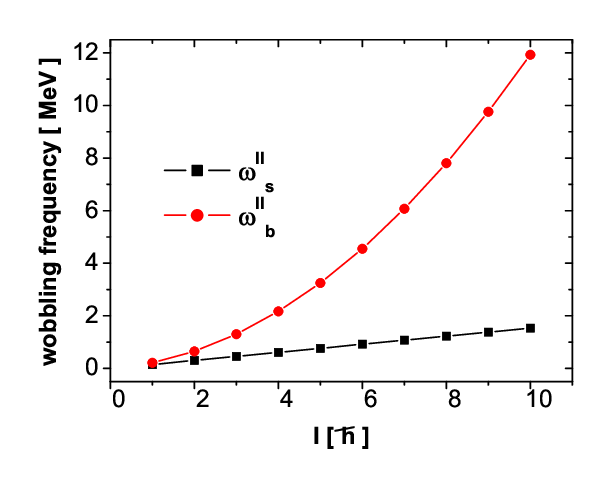}\includegraphics[height=5.5cm,width=0.5\textwidth]{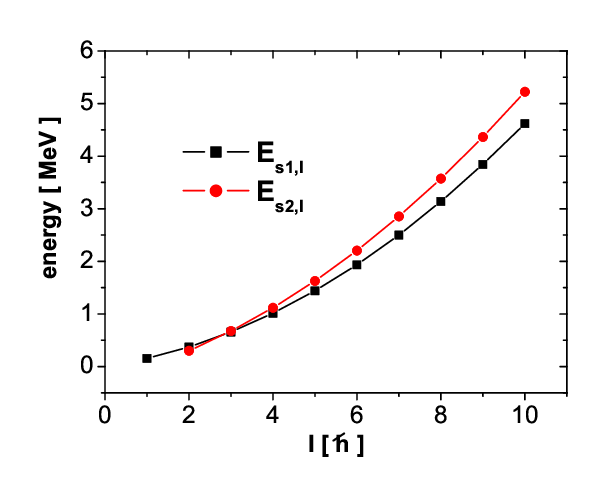}
\caption{Color online. Wobbling frequencies given by the semi-classical and boson expansion method, respectively are presented in the left panel. Energies for the yrast and the first one phonon excited bands corresponding to a semi-classical approach are shown in the right panel.}
\label{omenii}
\end{figure}
Using for the proton and neutrons moment of inertia the fraction $Z/A$ and $N/A$ of the global value from Eq.(\ref{inert}) and V=0.101 MeV (for this value the energies of the states $2^+$ from the yrast and the first one phonon excited band are different by an amount of about 0.1 MeV) we calculated the wobbling frequencies and the energies in the first two bands. The results are represented in Fig.\ref{omenii} as function of the total angular momentum. One notices that the two mentioned bands are crossing each other for $I$ equal to four. We don't show here the results for energies provided by the boson expansion method. The reason is that for this case energies are very large  for high angular momentum. This feature is caused by the fact that contrary to the semi-classical minimal energies shown in Fig.\ref{eminim}, in the boson picture the minimum energy is vanishing.

\begin{table}
\begin{tabular}{|c|cc|}
\hline
I &\multicolumn{2}{c|}{$B(M1,I^+  \to  (I+1)^+)[\mu_N^2$} \\
  & semi-classic& boson expansion\\
\hline
1&     0.029& 0.050\\
2&     0.057&0.154\\
3&      0.085& 0.315\\
4&     0.113&0.533\\
 5&    0.142&0.887\\
6&      0.1707& 1.1378\\
7&       0.198&1.506\\   
8&     0.227&1.969\\
9&     0.255&2.469\\
\hline
\end{tabular}
\caption{BM1 values for the inter-band transition $I^+\to (I+1)^+$ obtained within the semi-classical approach and boson expansion method.}
\end{table}
Using the same gyromagnetic factors as defined before we calculated the B(M1) values for  the inter-band magnetic dipole transitions within the framework of semi-classical and boson expansion formalisms respectively. We notice that the values corresponding to the boson picture are larger than those obtained semi-classically.

\begin{figure}[h!]
\includegraphics[height=5.5cm,width=0.5\textwidth]{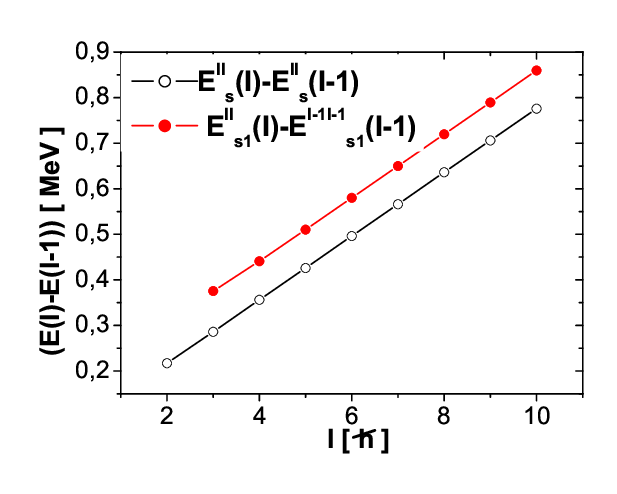}
\caption{Color online. Energy spacing in the ground and excited band.}
\end{figure}

\begin{figure}[h!]
\includegraphics[height=5.5cm,width=0.5\textwidth]{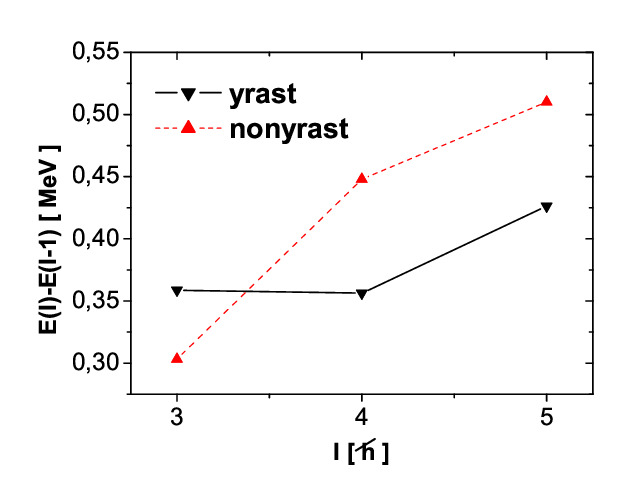}
\caption{Color online. Energy spacing in the yrast and non-yrast band.}
\end{figure}

In Ref.\cite{Cui} a projection shell model procedure was used to define a band built on top of the scissors mode at about 3 MeV, in $^{156}$Gd.
The band is characterized by a strong staggering effect. This effect reflects also in the relative values of the moment of inertia {MoI)} in the two $\Delta I =2$ branches of the $1^+$ band. Indeed the MoI for I=even is about 10 \% larger than that corresponding to odd spin.  This feature was identified also in  some isotopes and isotones of $^{156}$Gd. For all studied nuclei one finds  a lower than usual $2^+$ above the the scissors state. The mentioned effects are amplified by adding the two quasiparticle configurations. Although experimental data exist only for the first energy difference this is considered as a confirmation for the so called $\Delta I=2$ bifurcation effect. The staggering features shown by the microscopic description quoted above, were earlier predicted in Ref. \cite{RFC87}. In our case the states are different as nature, the ground band being built by adding the precession  and the zero point vibrations energies, while the excited band is obtained from the ground state band by adding the energy of the corresponding wobbling mode. Both are $\Delta I=1$ bands.  The excited band is headed by a $2^+$ state
which, as a matter of fact, is closer in energy than the state $2^+$ belonging to the ground band. As shown in Fig.5 none of the two bands shows a staggering effect. However, if a similar plot is performed for yrast and non-yrast bands a soft staggering shows up in the low part of the yrast band but not in the non-yrast band (see Fig.6). The mentioned staggering is caused by the crossing of the two semi-classical bands shown in the right panel of Fig.4. Note that while in Ref.\cite{Cui}, by contrast,  there is only one band with two branches one with even and one with odd spins. Another difference between our approach and that of Ref\cite{Cui} is that while in the quoted reference the states are of scissors type located around 3 MeV, the states described in this paper are of shears kind and lying in the low part of the spectrum.

We may ask ourselves why did we prefer a phenomenological instead of the existent appealing microscopic methods. The answer is that we aimed at outlining
the possible existence of low energy scissors-like states. Indeed, the parameters are not fundamented by a microscopic structure but rather used a systematic analysis of the deformation, energy and BE2 value characterizing the state $2^+$ in the ground band, as concerns the moment of inertia,  and fitting the predicted energy difference between the levels $2^+$ from the first and second scissors bands to an ad-hoc value of 0.1MeV. In this respect the proposed approach exhibits a lack of consistency which as a matter of fact does not affect the drawn qualitative conclusions In contrast to this feature, the approach shows simplicity, elegance and an intuitive interpretation of the pointed out phenomenon.

Note that we used the name of wobbling for the states described herein. This name was invented by Bohr and Mottelson in Ref.\cite{Bohr} for the sates predicted by an asymmetric rigid rotor and consisting in a precession motion of the total angular momentum around the axes with the maximal moment of inertia and an oscillation
of the angular momentum projection on the OZ axes. By inspection of Eq. (6.2) the motion of the present system has a similar structure. Concerning the asymmetry of the rotor system this is also simulated by a more complex geometry. Indeed, a system of two axial symmetric rotors with different symmetry axes is equivalent to a body exhibiting an hexadecapole deformation. Thus our naming is consistent with that of the standard wobbling motion.

A special focus is put on the band (I,I) for which the total angular momentum I is a good quantum number. This feature contrasts the case of TRM where the scissor state
has not the angular momentum as a good quantum number.
 We found out that wobbling motion is not the  exclusive virtue of the triaxial rotor system but also for a two axial symmetric rotors.       

\section{Conclusions}
A two axial symmetric rotor Hamiltonian is alternatively treated  semi-classically, via a Time Dependent Variational Principle,  and by a Dyson boson expansion method.
In both cases the linearized equations of motion lead to dispersion equations for the wobbling frequencies. A ground band is built up with the states having the 
zero-point energies. The corresponding phonon operators are exciting it to a second band. The energies of the two bands are compared with each others. The phonon amplitudes are used to calculate the B(M1) values regarding the reduced probability for the transition from the state $I$ from the ground band to the state $I+1$ from the excited band. 
Calculations are performed using Z=64 and N=92, i.e. data for $^{156}$ Gd.
The states from the two bands have a shears character.  It is shown that fixing the moment of inertia by fitting the energy of the K=2 state $2^+_{\gamma}$ , the energy of  the $1^+$ from the ground band is 
not very far from the  energy of the scissors dipole state, $1^+$.
  of the scissors mode 
The ground band (I,1) is degenerate with the band (1,I) due to the p,n permutation symmetry. However there exists another symmetry which is broken by the interaction term, namely the chiral symmetry. Indeed, changing the sign of either  $\vec{I}_p$ or $\vec{I}_n$ one calculated the chiral partner band of the ground band in the chiral untransformed picture. The energy split between the partner bands is increasing with $I$. In particular, according to the present approach there must exist a chiral partner state for the scissors mode identified at 3.075 MeV. 

Summarizing the main results we noticed that the variational function for a fixed pair of $(I_p,I_n)$ is a linear combination of states with the total angular momentum as a good quantum number. In order to avoid the tedious procedure of projection after variation, we proposed two scenarios for the pair $(I_p,I_n)$  where the average total angular momentum is close to $I_p$: a) In the situation of $(I_p,1)$ the component $I_p$ is a rough approximation of $\langle I\rangle$; b) In the situation of $I_p=I_n$ the common value of $I_p$ and $I_n$ is an excellent approximation 
for the average value $\langle I\rangle$. Within the limits of such approximations the states of the bands defined in the previous sections could be labeled by the total angular momentum I.
  
In conclusion, the two-rotor Hamiltonian successfully captures both wobbling and chiral features of nuclear systems. Experimental data in this context  would provide valuable motivation for further investigations.

\section{Appendix A}
\renewcommand{\theequation}{A.\arabic{equation}}
\setcounter{equation}{0}
The expressions for the coefficients involved in the linearized equations of motion are as follows:
\begin{eqnarray}
A_{11}&=&VI_p^2\frac{\sqrt{t(2I_n-t)}}{\left(r(2I_p-r)\right)^{3/2}}\left|_{\begin{matrix}r=\stackrel{\circ}{r}\cr
                                                                                  t=\stackrel{\circ}{t}\end{matrix}}\right.,\nonumber\\
A_{12}&=&\frac{V}{2}\left[\frac{1}{4I_pI_n}-2\frac{(I_p-r)(I_n-t)}{\sqrt{rt(2I_p-r)(2I_n-t)}}\left|_{\begin{matrix}r=\stackrel{\circ}{r}\cr
                                                                                  t=\stackrel{\circ}{t}\end{matrix}}\right.\right],\nonumber\\
A_{22}&=&VI_n^2\frac{\sqrt{r(2I_p-r)}}{\left(t(2I_n-t)\right)^{3/2}}\left|_{\begin{matrix}r=\stackrel{\circ}{r}\cr
                                                                                  t=\stackrel{\circ}{t}\end{matrix}}\right.,\nonumber\\
B_{11}&=&V\sqrt{rt(2I_p-r)(2I_n-t)}\left|_{\begin{matrix}r=\stackrel{\circ}{r}\cr
                                                                                  t=\stackrel{\circ}{t}\end{matrix}}\right.
\end{eqnarray}


\begin{thebibliography}{99}

\bibitem{GR64} W. Greiner, Phys. Rev. Lett. \textbf{14}, 599 (1965); W. Greiner, Nucl. Phys. \textbf{80}, 417 (1966); A. Faessler, W. Greiner, Z. Phys. \textbf{179}, 343 (1964)
\bibitem{FA66} A. Faessler, Nucl. Phys. \textbf{85}, 417 (1966)
\bibitem{MGM75} V. Maruhn-Rezwani, W. Greiner, J.A. Maruhn, Phys. Lett \textbf{57}, 109 (1975)
\bibitem{GKP70} S.I. Gabrakov, A.A. Kuliev, N.I. Pyatov, Yad. Fiz. \textbf{12}, 82 (1970); (Sov. J. Nucl. Phys. \textbf{12}, 44 (1971)); S.I. Gabrakov, A.A. Kuliev, N.I. Pyatov, D.I. Salamov, H. Schulz, Nucl. Phys. A \textbf{182}, 625 (1972)
\bibitem{KP74} A.A. Kuliev, N.I. Pyatov, Yad. Fiz. \textbf{20}, 297 (1974)
\bibitem{SR77} T. Suzuki, D.J. Rowe, Nucl. Phys. A \textbf{289}, 461 (1977)
\bibitem{IPF84} N. Lo Iudice, F. Palumbo, Phys. Rev. Lett. \textbf{74}, 1046 (1978); G. De Franceschi, F. Palumbo, N. Lo Iudice, Phys. Rev. C \textbf{29}, 1496 (1984)
\bibitem{BRS83} D. Bohle, A. Richter, W. Steffen, A.E.L. Dieprink, N. Lo Iudice, F. Palumbo, O. Scholten, Phys. Lett. B \textbf{137}, 27 (1983)
\bibitem{BBD84} U.E.P. Berg, C. Blassing, J. Drexler, R.D. Heil, U. Kneissle, W. Naatz, H. Ratzek, S. Schennach, R. Stock, T. Weber, H. Wickert, B. Fusher, H. Hollik, D. Kollewe, Phys. Lett. B \textbf{149}, 59 (1984)
\bibitem{IA81} F. Iachello, Nucl. Phys. \textbf{A358}, 89c (1981)
\bibitem{RFC87}A. A. Raduta, Amand Faessler and V. Ceausescu, Phys. Rev. C{\bf 36}, 2111 (1987).
\bibitem{IRD94}N. Lo Iudice, A. A. Raduta and D. S. Delion,Phys. Rev. C 50, 127( 1994).
\bibitem{RI89}A. A. Raduta, N. Lo Iudice, Z. Fur Phys. A {\bf 334} 403 (1989).
\bibitem{RD90}A. A. Raduta and D. S. Delion, Nucl. Phys. A{\bf 513}, 11 (1990).
\bibitem{Pena}D.Pena and Peter Ring,e-Print0912.0908 (nucl-th) 558-567.
\bibitem{Nest}V. O. Nesterenko, J. Kvasil, P. Vesely and P.G. Renhard, Int.J. Mod. Phys. E19 (2010)558-567.
\bibitem{Kroll}J. Kroll {\it et al.} European Physical Journal, DOI 10.1051/epjconf/20122104005.
\bibitem{RI90} A. Richter, Nucl. A. Phys. \textbf{507}, 99c (1990); A \textbf{522}, 139c (1991); A \textbf{553}, 417 (1993). Prog. Part. Nucl. Phys. \textbf{34}, 261 (1995)
\bibitem{IUD95} N. Lo Iudice, Prog. Part. Phys. \textbf{34}, 309 (1995)
\bibitem{ZAW98} D. Zawischa, J. Phys. G{\bf 24}, 683, (1998).
\bibitem{Rad017}A. A. Raduta, R. Poenaru and L. Gr. Ixaru, Phys. Rev. C {\bf 96}, 054320 (2017).
\bibitem{Ogu}T.Oguchi, Prog. Th. Phys.{\bf 26}, 26, 721 (1961).
\bibitem{Ring}P. Ring and P. Scuk, {it The Nuclear Many-Body Problem}, Springer- Verlag, Berlin, Heidelberg, New York, p. 19.
\bibitem{LRla}G. A. Lalazissis, S. Raman, P. Ring, Atomic Data and Nuclear Data Tables {\bf 71}, 1– 40 (1999).
\bibitem{Fin}M. Finger, T. I.Krakova, I Pricharzka and J. Ferenzei, Phys. Scr. 27,8 (1983).
\bibitem{Cui} Cui Juan Lv, Fang-Qi Chen. Yang Sun and Mike Guidry, Phys. Rev. Lett. 129,042502 (2022).
\bibitem{Bohr}A. Bohr and B. Mottelson, {\it Nuclear Structure} (Benjamin, Reading, MA, 1975), Vol. II, Ch. 4.
\end{thebibliography}
\end{document}